\documentclass[aps,prl,amsmath,amssymb,nofootinbib,superscriptaddress,twocolumn]{revtex4-1}
\usepackage{graphicx}
\usepackage{amsmath,pgfmath,pgffor}
\usepackage[usenames, dvipsnames]{color}
\usepackage{indentfirst}
\usepackage{hyperref}
\usepackage{subfigure}
\usepackage{epstopdf}

\usepackage{hyperref}

\newcommand{\be}{\begin{equation}}
\newcommand{\ee}{\end{equation}}

\usepackage{color}

\hypersetup{%
    pdfborder = {0 0 0},
    colorlinks,
    citecolor=blue,
    filecolor=Darkgreen,
    linkcolor=black,
}

\begin{document}

\title{Constraint satisfaction mechanisms 
for marginal stability \\ 
and criticality in large ecosystems}

\author{Ada Altieri} 
\affiliation{Laboratoire de Physique Th\'eorique de l'\'Ecole normale
sup\'erieure, PSL Research University, CNRS, Sorbonne Universités, 24 rue Lhomond, 75005 Paris, France;}

\author{Silvio Franz}
\affiliation{LPTMS, CNRS, Univ. Paris-Sud, Universit\'e Paris-Saclay, 91405 Orsay, France}

\begin{abstract}

We discuss a resource-competition model, which takes the MacArthur's model as a platform, to unveil interesting connections with glassy features and jamming in high dimension. This model, first studied in \cite{Tikhonov-Monasson}, presents two qualitatively different phases: a \emph{shielded} phase, where a collective and self-sustained behavior emerges, and a \emph{vulnerable} phase, where a small perturbation can destabilize the system and contribute to population extinction. 
We first present our perspective based on a strong similarity with continuous constraint satisfaction problems in their \emph{convex} regime. Then, we discuss the stability in terms of the computation of the leading eigenvalue of the Hessian matrix of the free energy in the replica space. This computation allows us to efficiently distinguish between the two aforementioned phases and to relate high-dimensional critical ecosystems to glassy phenomena in the low-temperature regime.

\end{abstract}

\maketitle

\emph{Introduction} - Critical properties of complex systems have been the object of an intense study in the last decades, both from a theoretical and an experimental perspective. 
On a theoretical level, those systems, which range from economy to biophysics, social sciences and game theory as well \cite{DeMartino_1, DeMartino_bio, DeMartino_bio2, Mora-Bialek, Mora}, are characterized by a large number of interacting components that might be responsible for collective behaviors. An increasing theoretical interest has also been fostered by a great revolution in microbial ecology \cite{Lozupone, micro, Chen}, which has allowed to unveil the complexity of microorganism communities. 

Despite the fact that single units themselves can follow very complex dynamical rules, statistical mechanics can be of help in the limit of infinite system size and well-mixed communities. On the one hand, the dynamical differential equations that govern ecological and biological populations are non-linear. As a consequence, their non-linearity might further complicate the analysis and generate limit cycles, as paradigmatically shown in the Lotka-Volterra-like models \cite{Volterra, Lotka, Bunin, Birolietal}.
Various approaches have been thus performed by taking into account either discrete or continuous growth, stochasticity in the demographic evolution as well as in the environment, symmetric or antisymmetric interactions, etc. 
One might ask, for instance, whether changing the connectivity, the degree of heterogeneity or the strength of the interactions between agents can either affect the stability of the system \cite{Yoshino} or generate chaotic regimes \cite{Bunin}. 
On the other hand, complex - and in particular living - systems generally display high susceptibilities. Using a physical metaphor, this corresponds to be poised at the edge of stability, namely very close to a special point in the parameter space: a so-called \emph{critical point} \cite{Kessler, Fisher, Mora-Bialek, Mora}. 

To give a clear-cut explanation of such phenomena, several mechanisms have been proposed: edge of chaos \cite{Kauffman}, self-organized criticality \cite{Bak}, self-organized instability, scale-free behavior and spontaneous symmetry breaking. 
From a complementary perspective, all these problems can also be interpreted as models of constrained satisfiability. Biological tissues and metabolic networks provide other examples in this direction \cite{DeMartino_bio, DeMartino_bio2, DDeMartino}, in which the optimization criterion relies on the minimization of a global function of reaction fluxes that corresponds to the decay rate of the entropy production in time. 
Very recent works suggest indeed that optimization in the presence of constraints \cite{SimplestJamming, FPUZ, Manning} can represent a natural mechanism for criticality and diverging susceptibilities. 

This intriguing mechanism has been recently deepened by studying the jamming phenomenon, a rigidity transition that occurs at zero temperature in amorphous systems with finite-range repulsive interactions \cite{Wyart_annales, Muller-Wyart, Liu-Nagel, Durian-Weitz, Liu-Nagel2010, CKPUZ17}. 
To clarify this point and deeper analyze the connections between the phenomenology of jamming and criticality in large ecosystems, we shall focus on a variant of the original MacArthur's model \cite{MacArthur}, whose phase diagram has been recently established by Tikhonov and Monasson \cite{Tikhonov, Tikhonov-Monasson}. 
In this letter we aim at studying the critical properties of the different emergent phases and computing, in particular, the spectrum of small harmonic fluctuations of the associated Lyapunov function. We shall also determine the leading eigenvalue of the stability matrix showing that its expression goes to zero upon approaching the transition between two qualitatively different phases and it can be thus associated with a diverging susceptibility. 

The novelty of our work mainly relies on the investigation of \emph{hidden} connections with constraint satisfaction problems and their mutual closeness to the low-temperature glassy physics, which, to the best of our knowledge, has not been investigated yet. 

\emph{Definition of the model} - The MacArthur's model describes an ecosystem with $S$ different species that compete for $N$ resources. It was originally studied for small ecosystems with a few species and resources \cite{MacArthur, Tilman} and only recently generalized to high dimension with novel, interesting collective behaviors. The number of individuals, $n_\mu$, where the index $\mu=1,..., S$ denotes the different species, depends on the availability of resources, $h_i$ ($i=1,..., N$), according to a feedback loop that is modulated by the efficiencies by which species exploit resources. 

One defines a total demand $T_i=\sum_\mu n_\mu \sigma_{\mu i}$ where the $\sigma_{\mu i}$'s are the \emph{metabolic strategies}, through which species demand and possibly meet their requirement.
The resource availabilities are thus chosen as a decreasing function of the total demand $T_i$, \emph{i.e.} $h_i= \frac{R_i}{\sum_\mu n_{\mu} \sigma_{\mu i}}$,
where $R_i$ is the resource surplus. The average value of the resource surplus is assumed to be constant and equal to one, whereas its variance, $\delta R^2$, represents a control parameter of the model. 
For each species $\mu$, the strategy is a random binary vector whose components $\sigma_{\mu i}$ are extracted from a distribution that takes values $1$ and $0$ with probabilities $p$ and $1-p$ respectively. The parameter $p$ determines whether the species in the ecosystem are either specialists ($p \ll 1$) - then each requires a small number of well-defined metabolites necessary to its survival - or generalists ($p \sim 1$), meaning that many different metabolites supply their needs. 

One can then introduce the \emph{resource surplus}, $\Delta_\mu$:
\begin{equation}
\Delta_\mu=\sum_{i=1}^{N} \sigma_{\mu i} h_i -\chi_\mu \ ,
\label{Delta}
\end{equation}
where $\chi_\mu$ stands for the species requirement. The above expression can be nevertheless rewritten more conveniently in terms of a random cost, \emph{i.e.} $\chi_\mu= \sum_{i} \sigma_{\mu i} + \epsilon x_\mu$, where $\epsilon$ is an infinitesimal formal parameter and $x_\mu$ is a unit variance Gaussian variable. The control parameters of the model can be eventually summarized in: $\alpha=S/N$, the \emph{density} of the species' pool; $\epsilon$, the width of the cost distribution; $p$, the average of the metabolic strategy distribution and $\overline{\delta R^2}$, the variance of the resource supply. 
Hence, depending then on the specific value of the resource surplus, $\Delta_\mu$, such a model can be interpreted as a specific instance of constraint satisfaction problem (CSP) \cite{Gardner, Gardner-Derrida, Engel}. Above the hyperplane $\vec{h} \cdot \vec{\sigma}_\mu$, species are able to survive and multiply; conversely, if $\vec{h} \cdot \vec{\sigma_{\mu}} < \chi_\mu$, the sustainability of the species' pool is not guaranteed anymore.
All $\vec{h}$ such that $\vec{h} \cdot \vec{\sigma_\mu} < \chi_\mu$ define the so-called \emph{unsustainable region}, for each species $\mu$.

\emph{Previous results} - The dynamics of the model is defined by the following differential equation:
\begin{equation}
\frac{d n_\mu}{dt} \propto n_\mu \Delta_\mu \ ,
\label{dn-mu}
\end{equation}
and, as far as one is concerned with equilibrium, any proportionality factor can be safely neglected. The condition for which Eq. (\ref{dn-mu}) vanishes leads only to two possibilities: i) $n_\mu >0$ $\&$ $\Delta_\mu=0$ (survival); ii)  $n_\mu =0$ $\&$ $\Delta_\mu<0$ (extinction). The case $\Delta_\mu >0$ is actually forbidden according to the definition of the model. For more details we refer the reader to \cite{Tikhonov-Monasson}. 

At equilibrium, the resource surplus distribution is \cite{Tikhonov-Monasson, DeMartino_1}:
\begin{equation}
p(\Delta)= \frac{1}{\sqrt{2\pi \psi^2}} e^{-\frac{(\Delta+p m)^2}{2 \psi^2}} \theta(-\Delta)+ H\left( \frac{p m}{\psi} \right) \delta(\Delta) 
\end{equation}
where the term $pm$ comes from the average over $\mu$, with a negative sign, while $H(x) \equiv \int_{x}^{\infty} \frac{dy}{\sqrt{2\pi}}e^{-\frac{y^2}{2}}$.
The $\delta$-contribution accounts for the fraction of species that actually satisfy the constraint and meet their requirement, while the second part is given by a shifted Gaussian distribution with standard deviation $\psi=\sqrt{p(1-p)q+\epsilon^2}$. The additional parameter $q$ denotes the relative variation of the resource avalabilites, \emph{i.e.} $q= \sum_i (1-h_i)^2$.

To better investigate the stability of the system against perturbations and invasions, one can moreover define a Lyapunov function \cite{Lyapunov, Tikhonov-Monasson}, increasing on each trajectory:
\begin{equation}
F(\lbrace n_\mu \rbrace)=\sum_{i} R_i \log \left(\sum_\mu n_\mu \sigma_{\mu i} \right) -\sum_\mu n_\mu \chi_\mu \ .
\label{lyapunov}
\end{equation}
Then, one can look at the partition function of the model, which can be eventually rewritten as:
\begin{equation}
Z \propto \int_{\Gamma} e^{-\beta \tilde{F}(\vec{h})} \prod_{\mu} \frac{1}{\beta( \chi_\mu-\vec{h} \cdot \vec{\sigma}_\mu)}
\end{equation}
where $\tilde{F}(\vec{h})$ denotes the Legendre transform of $F(\vec{n})$ and $\Gamma$ accounts for limiting the integration over the \emph{unsustainable region}.
Depending on the finite or divergent nature of the term $1/\Delta_\mu$, two main scenarios can take place. The differences are actually related to the finite-temperature regime or the zero-temperature limit: in the former, all species $n_\mu$ are characterized by a finite abundance, while in the second case only a finite subset is non-zero, which corresponds to those species with $\Delta_\mu$ exactly zero. This picture suggests that the system tends to rearrange in such a way to be poised at the edge of stability, namely to saturate the optimal equilibrium condition.  

Furthermore, it has been proven that, in the $\epsilon \rightarrow 0$ limit, the model undergoes a phase transition between two qualitatively different regimes, namely between a \emph{shielded phase} and a \emph{vulnerable phase} \cite{Tikhonov-Monasson}. In the shielded phase, $\mathcal{S}$, a collective behavior emerges with no influence of external conditions. If the availabilities are set to $1$, in such a way that neither specialists nor generalists are clearly favored, and a sufficiently small perturbation is applied to the system, a feedback mechanism between $h_i$ and $n_\mu$ contributes to adjusting mutual species' abundance and to keeping the availabilities almost unchanged, for all $i$. 
The situation is quite different in the \emph{vulnerable phase} phase, $\mathcal{V}$, where species cannot self-sustain and turn out to be strongly affected by changes and improvements in the immediate environment.

\emph{Marginal stability analysis} - Given the definition of the partition function, one can immediately extrapolate the asymptotic behavior at large $\beta$. In this limit, the free energy exhibits a logarithmic trend as a function of $\Delta_\mu$:
\begin{equation}
\log Z= \max_{\Gamma} \left( -\beta \tilde{F}(\vec{h}) -\sum_\mu \log \vert \Delta_\mu \vert \right) \ .
\label{Log_Z}
\end{equation} 
The logarithmic contribution becomes dominant in the low-temperature regime for which $\Delta_\mu$ tends to zero for a few selected species.
Remarkably, this result is reminiscent of the glassy phenomenology of hard-sphere systems and other continuous constraint satisfaction problems \cite{AFP, Altieri} close to the jamming transition, where a similar logarithmic behavior is consequence of a marginal mechanical stability.

Based on this observation, we then focus on the study of the density of fluctuations in both the $\mathcal{S}$ and $\mathcal{V}$ phases. Before entering into more detail, it is worth noticing that one of the most evident advantages of dealing with this well-mixed habitat model is the possibility to define a simple Lyapunov function, as in Eq. (\ref{lyapunov}).
The Lyapunov function, either of the availabilites or the abundances, is convex and bounded from above, and guarantees that an equilibrium state always exists. 
By differentiating the Lyapunov function to the second order we eventually obtain:
\begin{equation}
\frac{d ^2 F}{d n_\mu d n_\nu}=-\sum_{i} \sigma_{\mu i} \sigma_{\nu i} \frac{R_i}{(\sum_{\rho} n_{\rho} \sigma_{\rho i} )^2}=-\sum_{i} \sigma_{\mu i} \sigma_{\nu i} \left( \frac{h_i^2}{R_i} \right) .
\label{Hessian_n}
\end{equation}
In the $\mathcal{S}$ phase, \emph{i.e.} for $h_i \simeq 1$, the above matrix reduces to a Wishart matrix whose eigenvalue distribution is described by a Marchenko-Pastur law \cite{MarchenkoP} in the large-$N$ limit. Then, we can take advantage of the main results in random matrix theory to write the corresponding eigenvalue distribution as:
\begin{equation}
\rho(\lambda)= \frac{1}{2 \pi} \frac{\sqrt{(\lambda-\lambda_{-})(\lambda_{+}-\lambda)}}{\lambda} \ ,
\end{equation}
where the upper and lower edges of the spectrum are $\lambda_{\pm}= (\sqrt{[1]}\pm 1)^2$. We resort here to the same notation employed in disordered systems and CSP literature \cite{FPUZ, AFP}. In particular, the quantity $[1]$ multiplied by $N$ denotes the number of active species at criticality, or even the number \emph{sazieted constraints} for which $\Delta_\mu=0$. Because the lower edge of the spectrum $\lambda_{-} \rightarrow 0$ goes to zero at criticality \cite{DeMartino-Marsili, FPUZ}, as a signature of the approaching transition, the density contribution in the $\mathcal{S}$ phase simply reads:\begin{equation}
\rho(\lambda) \sim \sqrt{\left(4-\lambda\right)/{\lambda}} \ .
\end{equation} 
By contrast, in the $\mathcal{V}$ phase, where $h_i \neq 1$, in order to compute the spectrum of small harmonic fluctuations, one needs the distribution of availabilities $P(h_i)$ and their associated moments. 
Computing explicitly the expression of the availability distribution is a possibility. Another possibility is evaluating the eigenvalues of the stability matrix in the replica space. A usual trick to deal with disordered systems is indeed based on the replica formalism and the subsequent diagonalization of the Hessian matrix on a suitable subspace. 

Generically, the dependence on the replica indices can be expressed in terms of the projection on three different subspaces: the \emph{longitudinal} (scalar field), the \emph{anomalous} (one-replica-index) and the \emph{replicon} (tensorial field) \cite{DeAlmeida, DeDominicis-Giardina}. We focus on the replicon mode, as it is the most sensitive to the transition and responsible for possible replica symmetry broken solutions.
According to field-theory parlance, a zero (replicon) mode corresponds to a pole in the propagator and then a diverging \emph{spin-glass} susceptibility.
The longitudinal mode might instead give information about spinodal transitions, describing how a state opens up along an unstable direction and originates thus a saddle.

To simplify the computation, we can suppose to approach the transition from the so-called replica symmetric (RS) phase \cite{DeAlmeida}, where only a pure state exists, in agreement with the results by Tikhonov and Monasson. 
In this regime, the different correlators on which the replicon depends, can be expressed as appropriate combinations of error functions. The resulting expression for the replicon eigenvalue then reads:
\begin{widetext}
\be
\small
\lambda_{\text{repl}} \propto \int \mathcal{D}z  \frac{4 e^{-\frac{2(c+\psi z)^2}{b}} \left[ b+ 2 \sqrt{2 \pi b}(c+\psi z) e^{\frac{(c+\psi z)^2}{2b}} + 2\pi (c+\psi z)^2 e^{\frac{(c+\psi z)^2}{b} }  \right] }{b^3 \pi^2 \left[ 1+ \text{Erf} \left( \frac{c+\psi z}{ \sqrt{2 b}} \right) \right]^4 } \ ,
\label{integral_replicon}
\ee
\end{widetext}
where $\mathcal{D}z= \frac{dz}{\sqrt{2 \pi}} \; e^{-z^2/2}$ and the additional parameters are: $ b=p(1-p)(\overline{q}-q)$ and $c=pm^{*}$ (the label $^{*}$ means that a saddle-point computation has been taken over $m$). Given the definition of the overlap matrix $Q^{ab}= \frac{1}{N^2} \sum_{i}^{N} g_i^a g_i^b$, where $g_i$ is $N(1-h_i)$, we denote the difference between its diagonal and the off-diagonal values as $\overline{q}-q$. 
Note that in our computation the parameter $c$ is assumed to be positive. In principle, dangerous singularities might emerge for negative $c$ and a more detailed analysis would be needed.

Our central outcome relies on the fact that, upon approaching the $\mathcal{S}/\mathcal{V}$ transition, as $\epsilon \rightarrow 0$, the parameter $\psi$ tends to zero linearly. We show that in this regime a vanishing replicon mode appears, confirming the emergence of a marginally stable phase, which exactly coincides with the \emph{shielded} phase where $\psi=0$.  
We nevertheless remind that the sharp $\mathcal{S}/\mathcal{V}$ transition takes place for $\epsilon \rightarrow 0$ only, otherwise it would be replaced by a crossover regime.

\begin{figure}[h]
\centering
\includegraphics[scale=0.45]{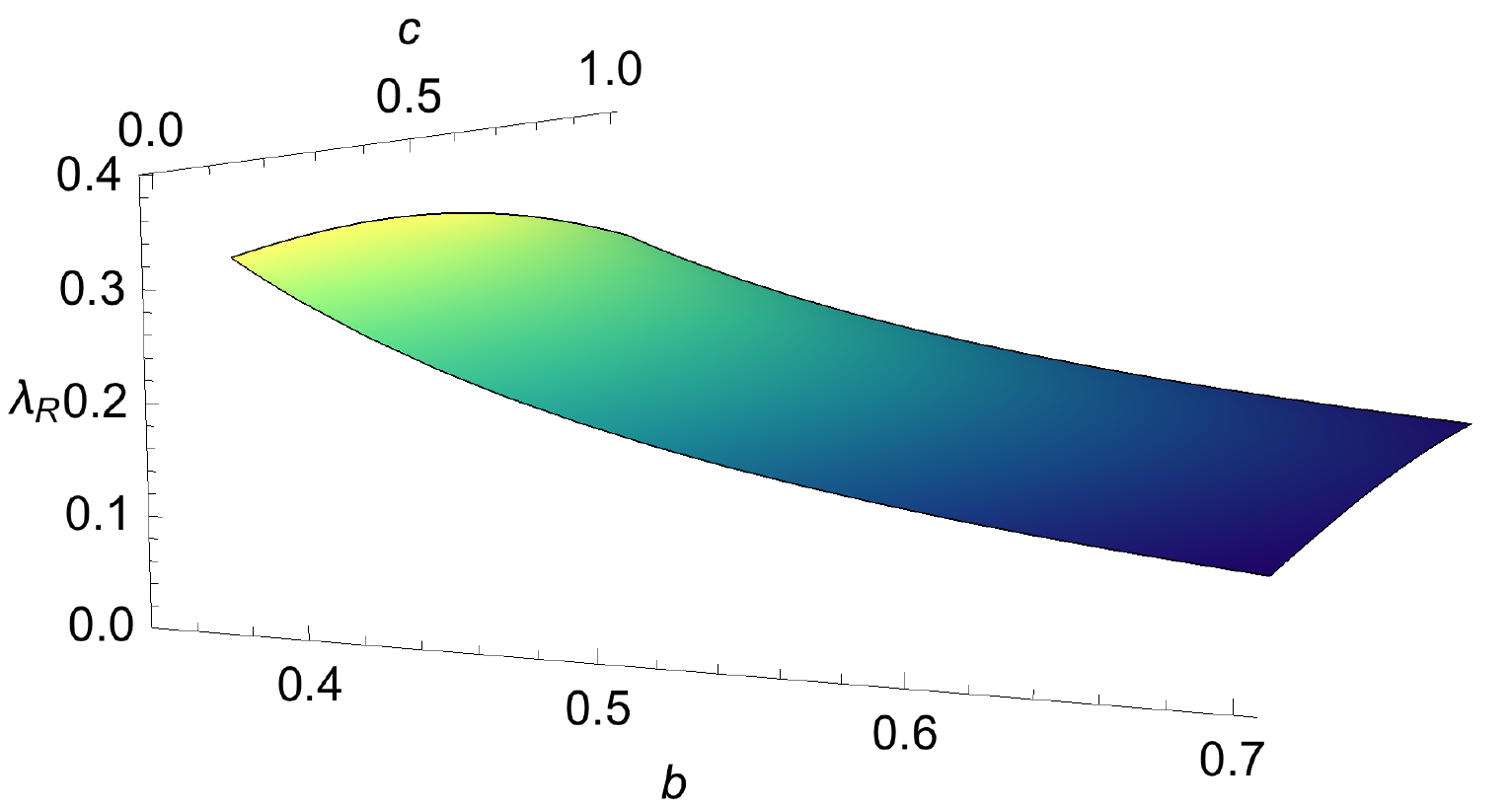}
\caption[$3D$ plot of the replicon eigenvalue]{Three-dimensional plot of the replicon eigenvalue at fixed and finite $\psi$, as a function of the parameters $b$ and $c$, as defined in the main text. In this regime, the replicon remains always finite.}
\label{3dplot_replicon}
\end{figure}

\begin{figure}[h]
\begin{minipage}{0.42\textwidth}
\includegraphics[width=\textwidth]{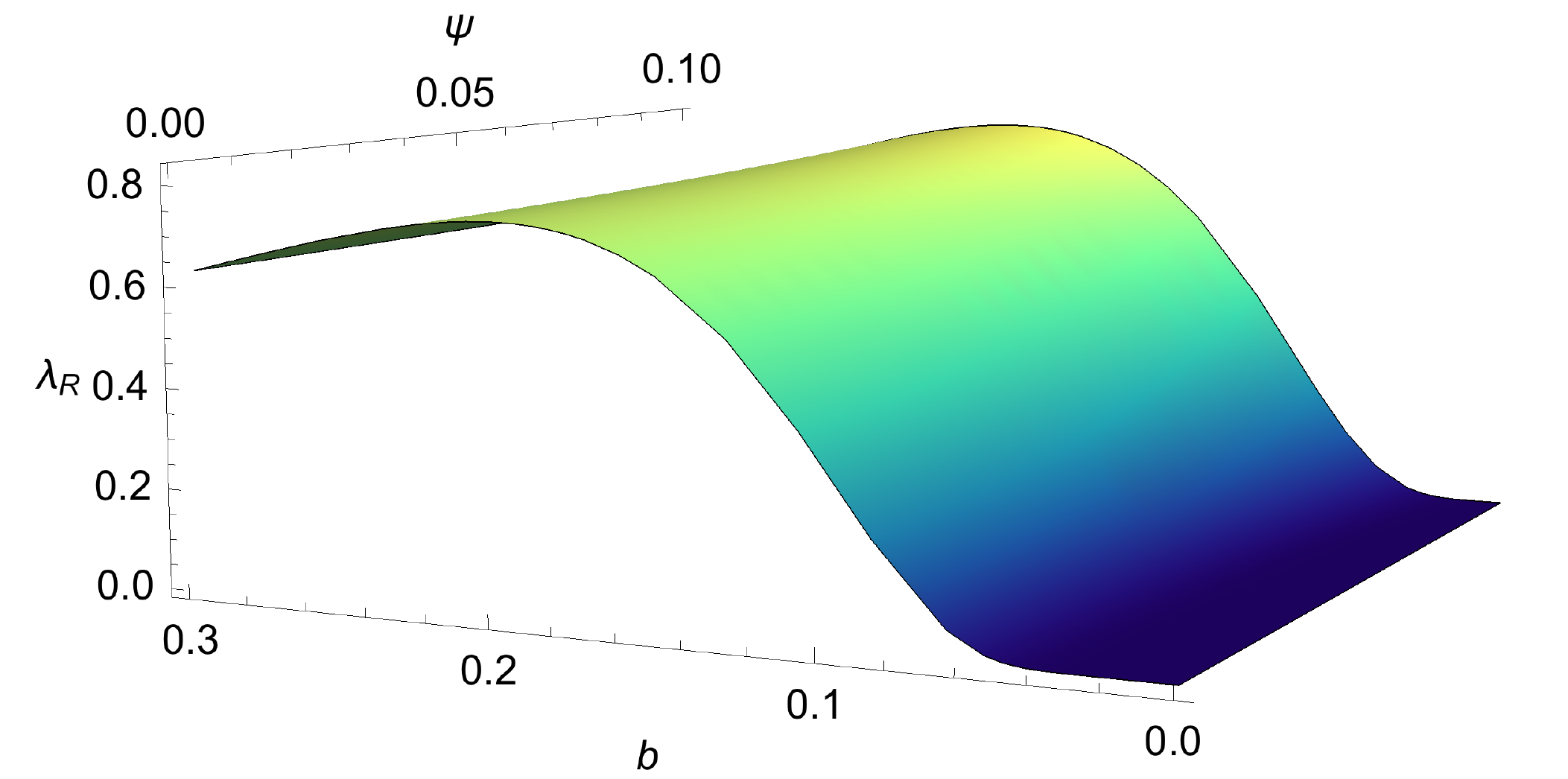}
\end{minipage}
\vspace{0.6cm}

\begin{minipage}{0.35\textwidth}
\includegraphics[width=\textwidth]{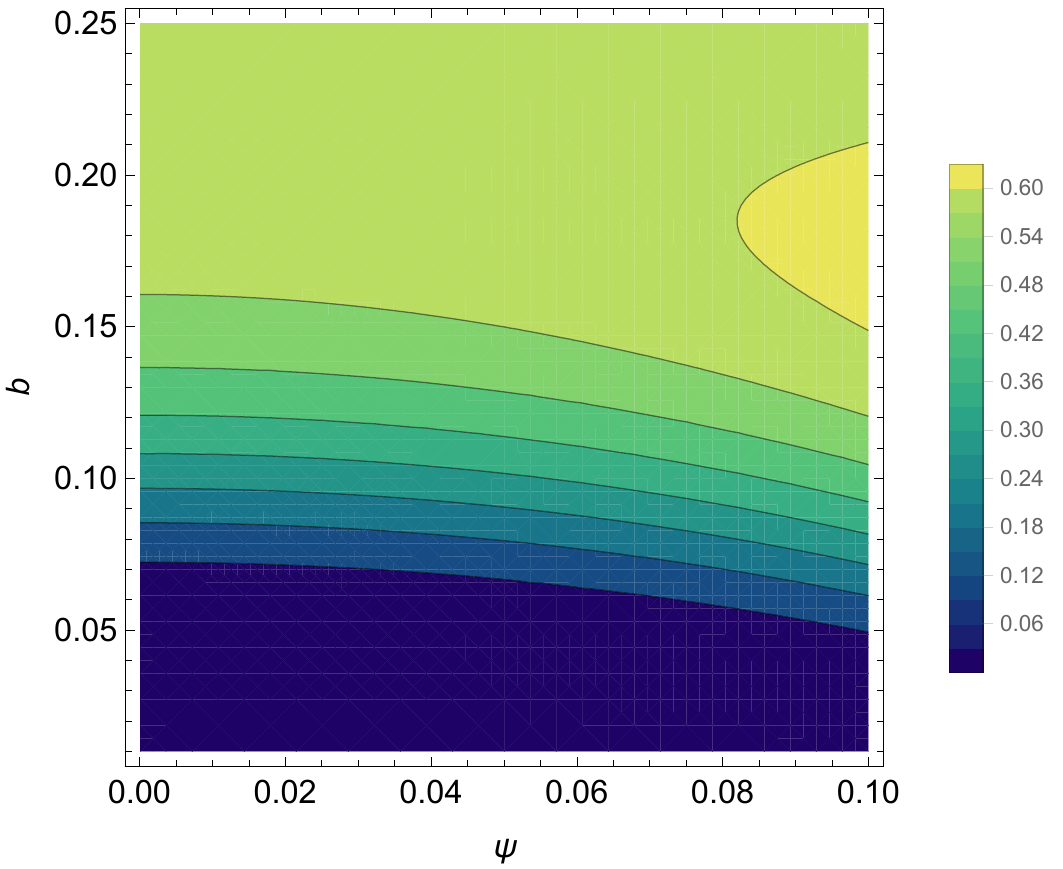}
\end{minipage}
\caption{On top, a three-dimensional plot of the numerical integration in Eq. (\ref{integral_replicon}). The solution is plotted as a function of $\psi$ and $b$ at fixed and positive $c$. The transition to the shielded phase is thus determined by a vanishing $\psi$, as $\epsilon \rightarrow 0$. In this regime, the replicon touches zero, as a signature of a marginally stable phase. Below, an isoline plot in the same parameter space at fixed $c$.}
\end{figure}
Hence, as soon as the system approaches the $\mathcal{S}/\mathcal{V}$ transition line, it self-adapts is such a way to keep its marginally stable nature. 
A large, diverging susceptibility - associated with the inverse of the replicon mode \cite{Mezard_Parisi_Virasoro, DeDominicis-Giardina} - is a further evidence that the system lies close to a critical point. As pointed out in recent works on collective behaviors in flocking birds \cite{Mora-Bialek}, a large response function can be interpreted as the fact that - rather than being governed by a single leader - the system tends to self-organize and to respond collectively to external perturbations. 

According to our interpretation, the $\mathcal{S}$ phase would correspond to an \emph{isostatic phase}, where the number of constraints equals the total dimension of the space and the spectrum becomes gapless. Conversely, the $\mathcal{V}$ phase would lead to a \emph{hypostatic regime}, with a number of constraints smaller than the threshold value \cite{Wyart_annales, SimplestJamming}. 
Our terminology is indeed based on the observation that, by increasing $\alpha$, the system enters the $\mathcal{S}$ phase where the number of \emph{active} species equals the number of resources $N$. The same picture has been exactly recovered by studying the jamming transition in analogous continuous constraint satisfaction problems \cite{SimplestJamming, FPUZ, AFP, Altieri}.

Supporting this scenario, a somehow complementary analysis was performed in an analogous model \cite{DeMartino-Marsili} governed by Lotka-Volterra-like equations in which the authors looked at the linear stability matrix and its smallest eigenvalue. However, compared to \cite{DeMartino-Marsili}, we have a different structure for the equations that modulate the resource abundances and a more involved expression for the resulting free energy. As a consequence, we focused on a different kind of susceptibility and then of stability matrix, explicitly computed in the replica space.

Note that the Lyapunov function defined thus far is convex everywhere and, according to that, a replica symmetry broken regime cannot occur. In fact, compared to other well-known optimization problems, the scenario that takes place here looks very close to a \emph{random linear programming problem}. Remarkably, even though replica symmetry continues to holds, a marginally stable regime arises for some specific values of the control parameter $\alpha$ \cite{SimplestJamming}.

\emph{Conclusions} - 
We studied a high-dimensional ecosystem problem, for which both the employed mathematical formalism and the physical interpretation behind suggest a clear mapping with low-temperature glassy behaviors. In particular, we managed to derive the asymptotic behavior of the spectral density, namely a gapless Marchenko-Pastur distribution, not yet highlighted for this kind of models. It appears as a powerful tool in order to distinguish between the \emph{shielded} and the \emph{vulnerable} phases. 
Moreover, from the exact computation of the Hessian matrix of the replicated action and its leading eigenvalue - which drops to zero upon approaching the $\mathcal{S}$ phase - we provided a clear evidence of an underlying criticality and the consequent diverging behavior of appropriate response functions. 

Our findings open the door to the study of several interdisciplinary problems and more specific aspects of the jamming transition. Indeed, the formalism used for generic constraint satisfaction problems (CSP) can be applied to a context of interacting agents, drawing not only a working strategy but also insights. 
We focused on a purely replica symmetric computation from which we nevertheless managed to make predictions about marginal phases. 
In any case, the possibility of extending our formalism beyond a convex-like version looks really challenging. 
We leave the investigation of possible breaking phases, signaled by an unstable, negative replicon mode, for future research. What a full RSB regime \cite{Mezard_Parisi_Virasoro} would specifically imply for biological and ecological settings with the appearance of a hierarchical structure of \emph{basins} in the rough energy/free-energy landscape - as shown for amorphous jammed systems \cite{CKPUZ_II, CKPUZ17} - is particularly fascinating and still an open matter of debate.

\vspace{0.3cm}

\emph{Acknowledgments} -
We thank R. Monasson and M. Tikhonov for insightful discussions about this work. 
We are also very grateful to G. Biroli, G. Parisi and F. Zamponi for their interest, and for suggesting new interesting directions. Finally, we thank E. Marinari and A. De Martino for very useful comments on the draft.  
This work was supported by a grant from the Simons Foundation ($\#$454941, Silvio Franz).

\end{document}